\documentclass[aps,prl,twocolumn,groupedaddress]{revtex4-2}

\usepackage[usenames,dvipsnames]{xcolor}
\usepackage{amsmath}
\usepackage{chemarr}
\usepackage{siunitx}
\usepackage{graphicx} 

\newcommand{\so}
   {\mathrel{\rlap{\raise1pt\hbox{$>$}}{\lower3pt\hbox{$\sim$}}}}
\newcommand{\io}
   {\mathrel{\rlap{\raise1pt\hbox{$<$}}{\lower3pt\hbox{$\sim$}}}}
\newcommand{\kt}{k_\text{B}T}

\begin{document}


\title{Field-mediated interactions of passive and conformation-active particles:\\multibody and retardation effects}


\author{Jean-Baptiste Fournier}
\affiliation{Universit\'e de Paris, CNRS, Laboratoire Mati\`ere et Syst\`emes Complexes (MSC), F-75013 Paris, France}


\date{\today}

\begin{abstract}
Particles in soft matter often interact through the deformation field they create, as in the ``cheerios" effect or  the curvature-mediated interactions of membrane proteins. Using a simple model for field-mediated interactions between passive particles, or  active particles that switch conformation randomly or synchronously, we derive generic results concerning multibody interactions, activity driven patterns, and retardation effects.
\end{abstract}


\maketitle

\section{Introduction}

Particles deforming a correlated elastic medium, by imposing  boundary conditions, undergo field-mediated interactions (in addition to direct van der Waals or electrostatic forces). At equilibrium, these mediated interactions come from the free energy that is stored in the deformation of the medium.
When this free-energy is dominated by  fluctuations, the mediated interaction is referred to as a Casimir interaction~\cite{Fisher:1978}. Otherwise, it is simply referred to as an elastically mediated interaction. The critical Casimir effect between colloids in a binary mixture in is an exemple of the former\cite{Hertlein:2008} and the popular Cheerios effect is an example of the latter~\cite{Vella:2005}.
Mediated interactions occur between interfaces, colloids, floating objects~\cite{Vella:2005} or proteins~\cite{Reynwar:2007} in soft-matter media such as critical binary mixtures~\cite{Hertlein:2008}, liquid crystals~\cite{Ajdari:1991,Poulin:1997}, capillary interfaces~\cite{Nicolson:1949,Hu:2005} and bio-membranes~\cite{Goulian:1993,Dan:1993,Dommersnes:1999,Bitbol:2012,Wel:2016}.

Mediated  interactions are non pairwise-additive\cite{Dommersnes:1999,Kim:2000,Fournier:2002,Yolcu:2014}. This comes from the fact that the field at the boundary of the particles is affected if a new particle is added in the range of the field correlation length. Another feature  of field-mediated interactions is that they may be subject to  retardation effects~\cite{Fournier:2014}. This is due to the relative slowness of the field, which is caused by the viscoelasticity of soft matter media, resulting in an overdamped diffusive behavior of the field. These effects could be particularly important when the particles embedded in the media are active particles, i.e., particles subject to an externally-imposed dynamics, such as artificial self-propelled particles~\cite{Deseigne:2010,Theurkauff:2012}, bacteria~\cite{Cates:2015}, molecular motors~\cite{Alvarado:2013}, and pumping~\cite{Prost:1996,Ramaswamy:2000} or multi-state particles such as proteins~\cite{Chen:2004}.

The examples listed above are all somewhat unique in that they involve different types of elasticity,  field dynamics and particle activity. In order to examine the consequences of the interplay between multibody interactions and retardation effects for both passive and active particles, we have studied the simplest model involving these features: a Gaussian elastic field with overdamped dynamics that mediates interactions among diffusing particles, which are either passive or undergoing conformational changes. Some properties of this model system, such as its phase diagram and the creation of non-equilibrium bands in the presence of activity, have been discussed in Ref.~\cite{Zakine:2018,Zakine:2020}. Here, we complement these results and we examine multibody and retardation effects, which we quantify and discuss in generic terms.

\section{Model}

We consider the simplest model that exhibits many-body field-mediated interactions between particles~\cite{Zakine:2018}. We take a Gaussian scalar field $\phi(\bm x)$ in two-dimensions (2D) to which pointlike particles are coupled quadratically.
The Hamiltonian for the field and the particles is given by
\begin{align}
\mathcal H = \int d^2x\left[\frac r2\phi^2+\frac12c(\bm\nabla\phi)^2\right]
+\sum_{k=1}^N\frac B2\left[\phi(\bm x_k)-S_k\phi_0\right]^2,
\label{eq:Hamiltonian}
\end{align}
where the variables $S_k=\pm 1$ are spin-like variables encoding the state, or conformation, of the particle. Particles with $S_k=1$ (resp.\ $S_k=-1$) will be represented in white (resp.\ black). The white particles favor $\phi(\bm x_k)=+\phi_0$ with a strength $B$ at their position $\bm x_k$, and the black particles symmetrically favor $\phi(\bm x_k)=-\phi_0$. The correlation length of the bare field is $\xi=\sqrt{c/r}$. 

For the time evolution of the field, we resort to a purely relaxational dynamics satisfying detailed balance:
\begin{align}
&\partial_t \phi(\bm x,t)=-\Gamma\frac{\delta H}{\delta \phi(\bm x,t)}+\sqrt{2\Gamma k_\text{B} T}\,\xi(\bm x,t),
\label{eq:field_dyn}\\
&\langle\xi(\bm x,t)\xi(\bm x', t')\rangle=\delta(\bm x-\bm x')\delta(t-t').
\label{eq:field_dyn2}
\end{align}
where $T$ is the temperature, $\Gamma$ the mobility and $\xi(\bm x,t)$ a Gaussian white noise.
For the diffusion of the particles we adopt equilibrium overdamped Langevin equations satisfying detailed balance:
\begin{align}
&\frac{d\bm r_k}{d t}=-\mu \frac{\partial H}{\partial \bm r_k}+\sqrt{2\mu k_\text{B}T}\bm\eta_k(t),
\label{eq:particles_dyn}
\\
&\langle\eta_{k,i}\eta_{k,j}\rangle=\delta_{kl}\delta_{ij}\delta(t-t').
\label{eq:particles_dyn2}
\end{align}
with $\mu$ the mobility of the particles and the $\bm \eta_k(t)$ independent Gaussian white noises.

Finally, for the dynamics of the state of the particles, we consider three possibilities:
\begin{enumerate}
\item Time-independent states $S_k$ (passive particles). With this choice the whole system is at equilibrium.
\item Random flips with fixed symmetric rates (active particles):
\begin{align}
S_k=\mathrm{-1}\, \xrightleftharpoons[~\alpha~]{~\alpha~} \,S_k=\mathrm{+1}.
\label{eq:spin_flips}
\end{align}
With this choice detailed balance is broken and the whole system is out-of-equilibrium.
\item Synchronized deterministic flips every $\Delta t$ (active particles):
\begin{align}
S_k\overset{\Delta t}{\longleftrightarrow}-S_k,
\label{eq:spin_synchro}
\end{align}
With this choice the system is also out-of-equilibrium.
\end{enumerate}
The nonequilibrium aspects of this model in case~2 were discussed in Refs.~\cite{Zakine:2018,Zakine:2020}.

In the following, we shall work in dimensionless units such that $k_\text{B}T=\Gamma=a=c=1$. Here $a$ is the real-space cutoff describing both the actual size of the poinlike particles and the lattice spacing in the numerical simulations. In other words, lengths are normalized by $a$, energies by $k_\text{B}T$, and times, such as the time step $\tau$ in the simulations, by $a^2/(\Gamma c)$, while $c$ is absorbed in a redefinition of $\phi$. We shall assume $r\ll1$ in order for the field correlation length $\xi=\sqrt{1/r}\gg1$ to remain much larger than the cutoff length and the (effective) size of the particles.

To investigate the dynamics of the field and the particles, and in particular the many-body effects, we resort to two different numerical simulations (details in Appendix~\ref{apx:numerical_simulations}). In simulation~$1$, which is a lattice-based discretization of the stochastic eqns~\eqref{eq:field_dyn}--\eqref{eq:field_dyn2} and~\eqref{eq:particles_dyn}--\eqref{eq:particles_dyn2}, the particles have no direct interactions, except for the (very weak) mutually excluded volume discussed below. However, they interact with the field, which mediates effective interactions between the particles. In simulation~$2$, the field is absent, and the particles interact through direct pairwise forces corresponding to the two-body component of the field-mediated interaction. We  use in this case a lattice-based discretization of the stochastic eqns~\eqref{eq:field_dyn}--\eqref{eq:field_dyn2} with $\partial H/\partial\bm r_k$ replaced by the pairwise interaction.

Unless otherwise specified, instead of taking periodic boundary conditions, we take hard walls and set $\phi=0$ on the boundary (which only affects the system on a layer of width $\xi$). This is advantageous because  we then have  only a finite number of particles interacting through direct forces in simulation~$2$. 

We set a weak excluded volume between the particles by allowing up to $f=5$ particles on the same site. In simulation~$1$, this is almost equivalent to having no excluded volume at all since the average occupation per occupied site will turn out to be of order $1$. This means that the system would not be sensitive to the detail of short-range repulsive interactions. In simulation~$2$, the weak excluded volume is necessary as the particles will turn out to strongly aggregate. We chose a large $f$ in order to keep a fluid simulation while still evidencing the formation of clusters.

\section{Multibody versus pairwise interactions (passive particles)}

In this section we assume that the particles do not change state (case 1). The whole system is then in equilibrium.

\subsection{Pairwise interaction}

The pairwise field-mediated interaction between two particles with states $S_1$ and $S_2$ that are separated by a distance $R$, is calculated in Appendix~\ref{apx:field-mediated_interactions}. It is given by
\begin{align}
F_\text{pw}(R)=-\phi_0^2S_1S_2\,\frac{G(R)}{\left(B^{-1}+G_0\right)
\left[B^{-1}+G_0+S_1S_2G(R)\right]
}.
\label{eq:pairwise}
\end{align}
with
\begin{align}
G(R)&=\frac1{2\pi}\text{K}_0(R\sqrt{r}),
\quad G_0=\frac1{2\pi}\ln\!\left(\frac\pi{\sqrt{r}}\right),
\end{align}
where $\text{K}_0$ is the  Bessel function. This expression does not contain the fluctuation-induced composant of the interaction, i.e., the Casimir-like force, which is negligible when $\phi_0$ is is sufficiently large (as assumed throughout). Note that with this interaction particles of like states attract each other while particles of opposite states repel each other. 

\begin{figure}
\centering
  \includegraphics[width=1\columnwidth]{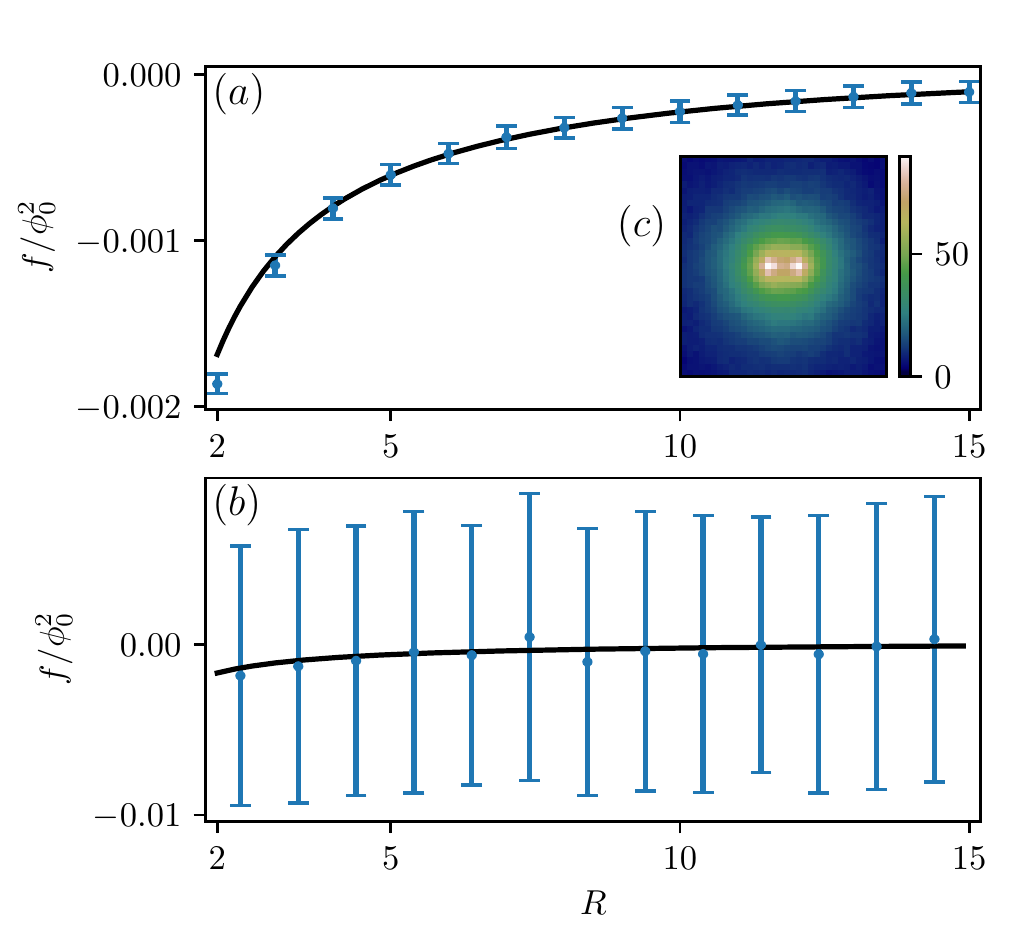}
  \caption{Field-mediate force $f$ between two identical particles normalized by $\phi_0^2$ as a function of their separation~$R$. Blue data: average value and standard deviation of the force computed from numerical simulation~$1$.  Solid black line: analytical force $f_\text{pw}(R)=-dF_\text{pw}/dR$ obtained from eqn~\eqref{eq:pairwise}. The parameters are $L=60$, $r=0.01$, $B=0.17$, $\tau=10^{-4}$ and (a) $\phi_0=1000$ or (b) $\phi_0=8$. (c) field map for a separation $R=5$ with the parameters of (a).
}
  \label{fgr:Force}
\end{figure}

Figure \ref{fgr:Force} shows  a comparison between the analytical force deriving from $F_\text{pw}(R)$ (calculated with field-theoretic methods in Appendix~\ref{apx:field-mediated_interactions}) and the stochastic force obtained during simulation 1 (see  eqn~\eqref{eq:discretized_force} in Appendix~\ref{apx:numerical_simulations}) for two particles  held at a fixed distance. In fig.~\ref{fgr:Force}a, where $\phi_0$ is very large, the agreement is well revealed as the fluctuation of the numerical force is small. The slight discrepancy at small distances is due to  finite size effects. In fig.~\ref{fgr:Force}b, where $\phi_0$ is smaller (but still much larger than unity) the agreement is still correct, but the standard deviation of the force is very large. In the following we shall use  values of $\phi_0\approx10$ comparable with that of fig.~\ref{fgr:Force}b.

\subsection{Multibody interactions}

As show in Appendix~\ref{apx:field-mediated_interactions}, the  field-mediated interaction between all the particles is nonpairwise. It is given by
\begin{align}
\label{eq:FMI}
F_\text{int}=\frac12\phi_0^2(S_1\ldots S_N){\mathsf A}^{-1}(S_1\ldots S_N)^t-
\frac12\,\frac{N\phi_0^2}{B^{-1}+G_0},
\end{align}
where $\mathsf A$ is a $N\times N$ matrix with diagonal elements $A_{ii}=B^{-1}+G_0$ and off-diagonal elements $A_{ij}=G(\bm x_i-\bm x_j)$. The second term in $F_\text{int}$ is the self-energy of the particles.

\begin{figure}
\centering
  \includegraphics[width=1\columnwidth]{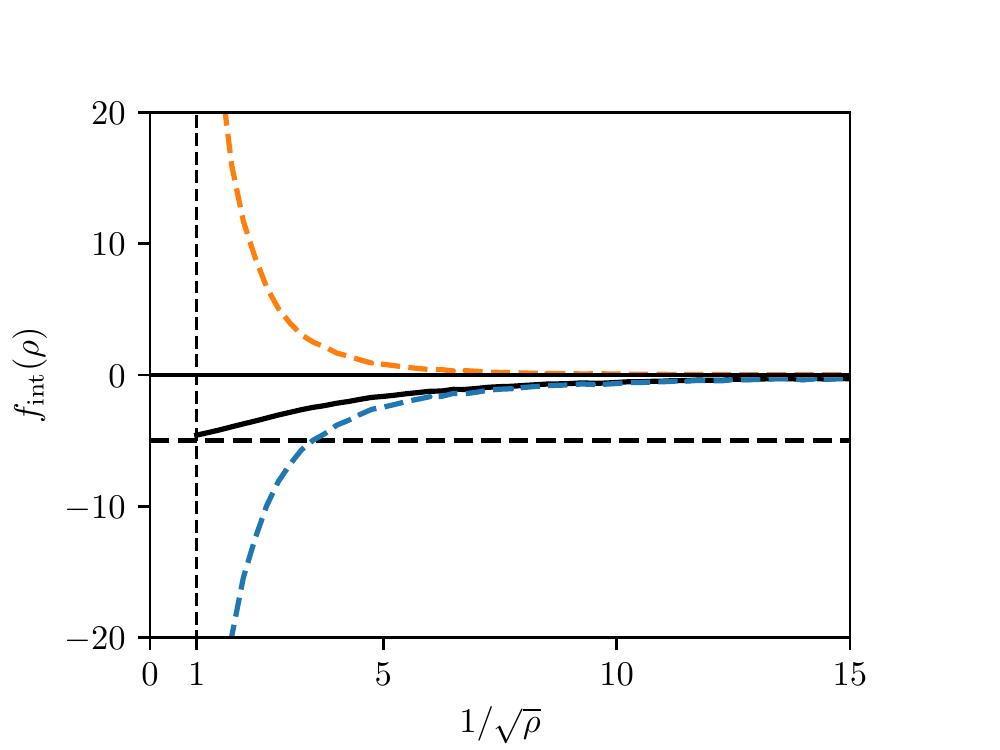}
  \caption{Solid black line: field-mediated interaction energy per particle $f_\text{int}(\rho)=F_\text{int}(\rho)/N$ obtained analytically (Appendix~\ref{apx:field-mediated_interactions}), for a 
a random distribution of identical particles, as a function of the inverse density $1/\sqrt{\rho}$ (typical interparticle distance). Dashed black line: minus the self-energy of an isolated particle, $-f_\text{self}$. Dashed blue line: pairwise interaction energy per particle, $F_\text{pw}/N$. Dahed orange line: multibody interaction energy per particle, $(F_\text{int}-F_\text{pw})/N$. Parameters: $r=0.01$ (i.e., $\xi=10$), $B=0.17$, $\phi_0=8$, $L=100$.}
  \label{fgr:Multibody_vs_pairwise}
\end{figure}

To study the importance of the multibody, i.e., non pairwise, contributions to the field-mediated interaction, we have calculated from eqn~\ref{eq:FMI} the total field mediated interaction energy in an assembly of identical particles (e.g., white ones, $S_k=1$). To do so, we take a $100\times100$ latice which we fill randomly with a density $\rho$ of particles.
We increased the density up to $\rho=1$, corresponding to a dense assembly of $N=10^4$ particles. Figure~\ref{fgr:Multibody_vs_pairwise} shows the field-mediated energy, the pairwise energy and the multiboly energy per particle as a function of the inverse density $1/\sqrt{\rho}$ (typical interparticle distance). It shows that
\begin{itemize}
\item The particles effectively interact only when they are separated by distances $\io\xi$.
\item The total field-mediated interaction (solid black line) is much weaker than its pairwise contribution (dashed blue line).
\item Multibody effects are repulsive and very strong (dashed orange line).
\item In a dense assembly the field-mediated interaction essentially relaxes the self-energy (see the solid and dashed black lines almost merging at $\rho=1$).
\end{itemize}

The physics can be understood as follows. In a dense assembly of particles, the field is almost uniform and close to $\phi_0$ (for large couplings). This implies that adding a particle is almost costless. So the interaction per particle is approximatively minus the self-energy: the energy relaxed by bringing the particle in the assembly. More quantitatively, the field-mediated energy per particle is approximatively given, for $\rho\approx1$, by
\begin{align}
f_\text{full}\simeq\min_\phi\left(\frac12r\phi^2+\frac12B(\phi-\phi_0)^2\right)=\frac12\frac{\phi_0^2}{B^{-1}+r^{-1}}.
\end{align}
Subtracting $f_\text{self}$, given by eqn~\eqref{eq:f_self} (see Appendix~\ref{apx:field-mediated_interactions}), we obtain the interaction energy per particle in a dense assembly:
\begin{align}
\bar f_\text{int}=
f_\text{full}-f_\text{self}\simeq\frac12\phi_0^2\frac{G_0-r^{-1}}{\left(B^{-1}+r^{-1}\right)\left(B^{-1}+G_0\right)}.
\end{align}
With $r=0.01$, $B=0.17$ and $\phi_0=8$ (same values as in Ref~\cite{Zakine:2018}), this gives $\bar f_\text{int}\simeq
-4.7$ in agreement with Fig.~\ref{fgr:Multibody_vs_pairwise}. By varying $B$ over more than two decades for $r=0.01$ and $r=0.1$, we have checked that the above formula agrees very well with exact multibody analytical calculations. Since $f_\text{self}=\frac12\phi_0^2/(B^{-1}+G_0)$, with $G_0$ of order unity, it is straightforward to see that $f_\text{full}\ll |f_\text{self}|$ when $B\gg r$, i.e., that the field-mediated interaction  effectively relaxes the self-energy. 

Conversely, the pairwise interaction per particle in a dense assembly, $\bar f_\text{pw}$ for $\rho\simeq1$, is extremely large. Indeed, it is given by
\begin{align}
\bar f_\text{pw}\simeq\int_1^\infty 2\pi R\,dR\, F_\text{pw}(R)\approx
-s_1s_2\phi_0^2\frac{\text{K}_1(\sqrt{r})}{\sqrt{r}\left(B^{-1}+G_0\right)}.
\label{eq:fpw_dense}
\end{align}
The above analytical result, obtained by neglecting the term $s_1s_2G(R)$ in the denominator of eqn~\eqref{eq:Fpw} (see Appendix~\ref{apx:field-mediated_interactions}), is valid when $B\io 1$. For $r=0.01$, $B=0.17$ and $\phi_0=8$, we obtain $\bar f_\text{pw}\simeq-150$ in good agreement with the exact calculation in a dense assembly. This is much larger than $\bar f_\text{int}\simeq-4.67$ obtained above. The pairwise interaction is therefore a very poor approximation of the field-mediated interaction in a dense system.

\begin{figure}
\centering
  \includegraphics[width=1\columnwidth]{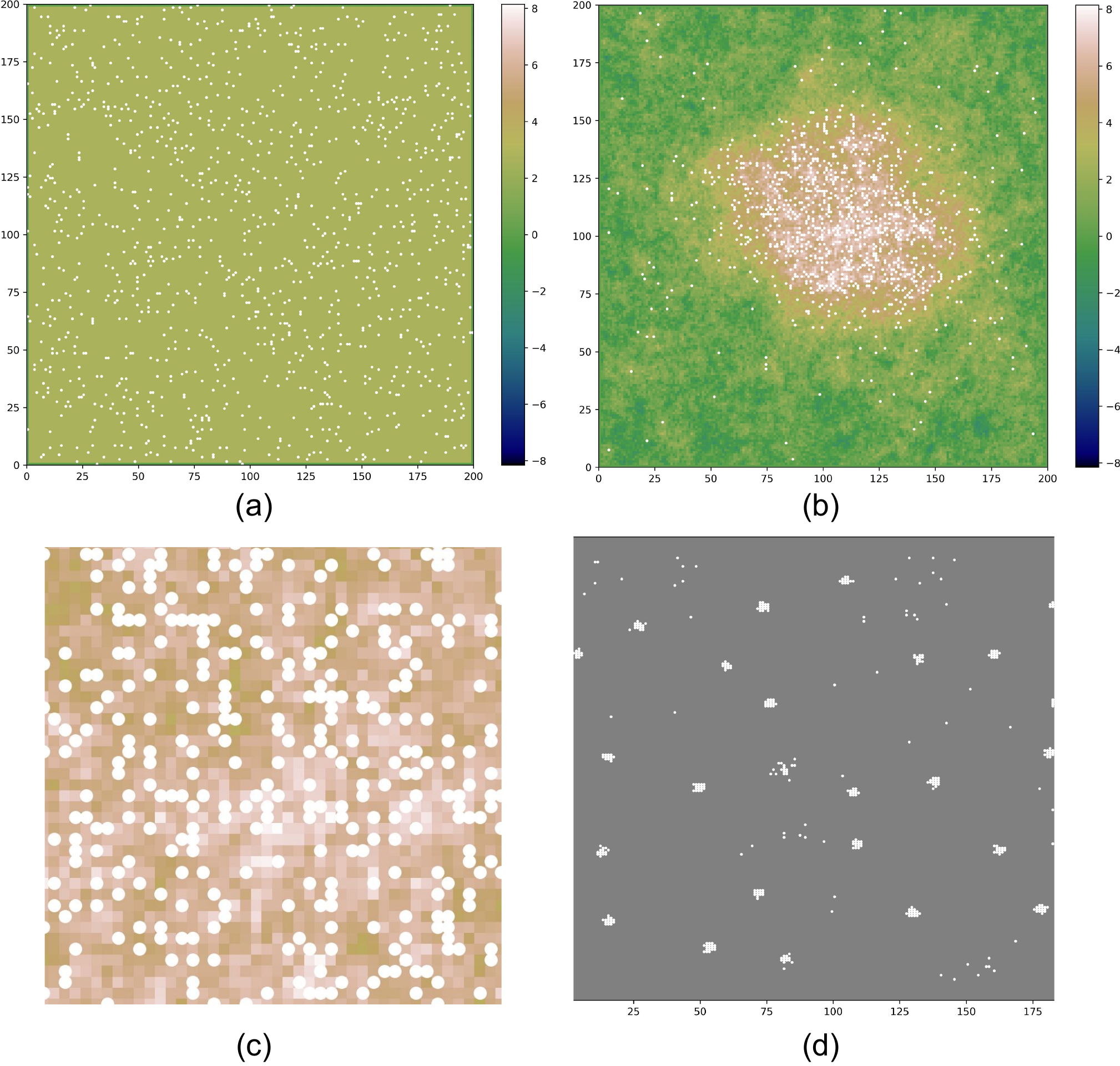}
  \caption{(a-b) Liquid-gaz phase separation arising from the field-mediated interactions between passive particles. The particles are the white dots and the color map indicates the field. Hard walls prescribe a zero field on the boundary. (a) Initial state at $t=0$. (b) Spontaneous phase separation into a liquid phase of density $\rho_l\approx0.2$ and a gaz phase of density $\rho_g\approx0.003$, at $t=3000$. (c) Zoom in the center of the liquid phase. (d) Same system, with same initial condition, but the particles interact directly with the pairwise interactions equal to the two-body component of the mediated interaction, at $t=63$. Dense clusters form that slowly merge (some still in formation). Parameters: $L=200$, $\rho=0.03$, $r=0.01$, $B=0.17$, $\phi_0=8$, $\mu=10$, $\tau=10^{-4}$. Average occupation $o$ per occupied site: (a) $o=1$. (b) $o=1.1$. (d) $o=2.7$ (rising).}
  \label{fgr:LG}
\end{figure}

\subsection{Liquid--gaz phase separation of identical  particles}

In order to investigate the collective behavior of the particles and observe the effects of multibody interactions, we have studied a system of $N$ identical particles using simulations 1 and 2 (fig.~\ref{fgr:LG}).

In simulation 1, with the same numerical parameters as in the previous subsection, we observe a liquid--gaz phase separation (figs.~\ref{fgr:LG}a and~\ref{fgr:LG}b). In the liquid phase, the field is roughly homogoneous, the equilibrium particle density is significantly less than unity ($\rho_l\simeq0.2$) and it is largely fluctuating. The average occupation per occupied site is $o\simeq1.1$ which implies that short-range repulsive forces are not involved in the stability of the liquid phase (recall that we allow up to $f=5$ particles to occupy the same site). This contrasts with ordinary liquid whose density is mainly controlled by the short-range repulsion between the molecules.

In simulation 2, we investigate the same system, but the field-mediated interactions, that include all multibody terms, is replaced by its pairwise component $F_\text{pw}$ only. The difference is striking: the system builds dense clusters stabilized by the excluded volume (figs.~\ref{fgr:LG}c and~\ref{fgr:LG}d). The comparison between the two simulations reveals thus the very strong countereffect of the multibody interactions, in agreement with the results of the previous subsection.

\begin{figure}
\centering
  \includegraphics[width=1\columnwidth]{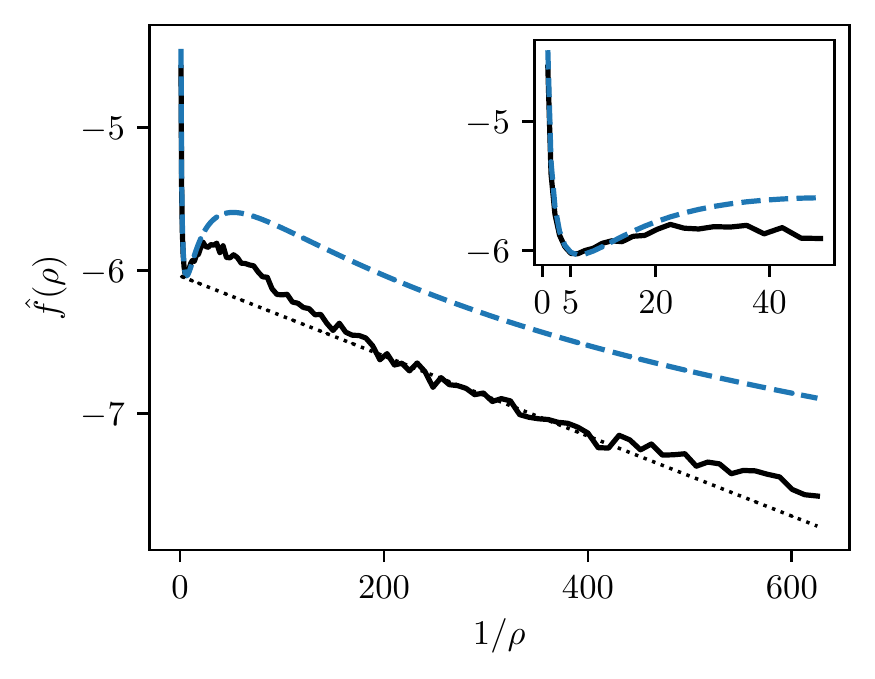}
  \caption{Double tangent construction revealing the liquid--gaz phase separation of passive particles subject to field-mediated interactions. Solid black curve: free energy per particle $\hat f(\rho)$, given by \eqref{eq:hatfrho}, as a function of the area per particle $1/\rho$. It is derived from the solid black line of fig.~\ref{fgr:Multibody_vs_pairwise} (where the noise is less visible due to the scale) by adding the entropy of mixing. Dotted line: double tangent. Dashed blue line: mean-field approximation $\hat f_\text{mf}(\rho)$ as given by \eqref{eq:mf}. Inset: zoom around the minimum. Parameters: same as in fig.~\ref{fgr:Multibody_vs_pairwise}.}
  \label{fgr:DT}
\end{figure}

To check if we can recover this behavior from the field-mediated interactions calculated analytically in Appendix~\ref{apx:field-mediated_interactions}, we consider the free energy per particle and we add the entropy of mixing:
\begin{align}
\hat f(\rho)=\frac1N F_\text{int}(\rho)+\ln\rho+(\rho^{-1}-1)\ln(1-\rho).
\label{eq:hatfrho}
\end{align}
As shown in fig.~\ref{fgr:DT} the double tangent construction predicts very well the phase separation, giving accurately the densities of the liquid and the gaz phases: $1/\rho_l\simeq5$ and $1/\rho_g\simeq300$.

\subsubsection*{Mean-field density functional approach}

In such systems, it is customary to use a density functional formulation then a mean-field approximations~\cite{Zakine:2018, Zakine:2020,Goutaland:2021}. Let us test the quality of this approach. The density functional associated to the Hamiltonian~\eqref{eq:Hamiltonian} for a system of identical (white) particles is
\begin{align}
F[\phi,\rho]=\int\!d^2x&\Big[\frac r2\phi^2+\frac12\left(\bm\nabla\phi\right)^2+\frac B2\rho\left(\phi-\phi_0\right)^2
\nonumber\\
&+\rho\ln\rho+\left(1-\rho\right)\ln\left(1-\rho\right)\Big]
\end{align}
In the mean-field approximation, where spatial variations are disregarded, the associated free energy per particle is thus given by
\begin{align}
\hat f_\text{mf}(\rho)&=\min_\phi\left(\frac r{2\rho}\phi^2+\frac B2\left(\phi-\phi_0\right)^2\right)
\nonumber\\&
+\ln\rho+(\rho^{-1}-1)\ln(1-\rho),
\label{eq:mf}
\end{align}
It is clear form Fig.~\ref{fgr:DT} that the double tangent construction based on the corresponding dashed blue line would predict quite well the liquid phase, but it would fail predicting correctly the density of the gaz phase, in which the particles are separated on average by a distance larger than $\xi$. Indeed, in the gaz phase, the density functional approach underestimates significantly the free energy per particle and overestimates at least by a factor of two the density of the gaz phase. Clearly, this approach works well in the liquid phase because the field is almost uniform there. As expected from the discussion in the previous subsection, it does capture well the multibody interactions there.

\subsection{Phase separation of opposite particles}

\begin{figure}
\centering
  \includegraphics[width=1\columnwidth]{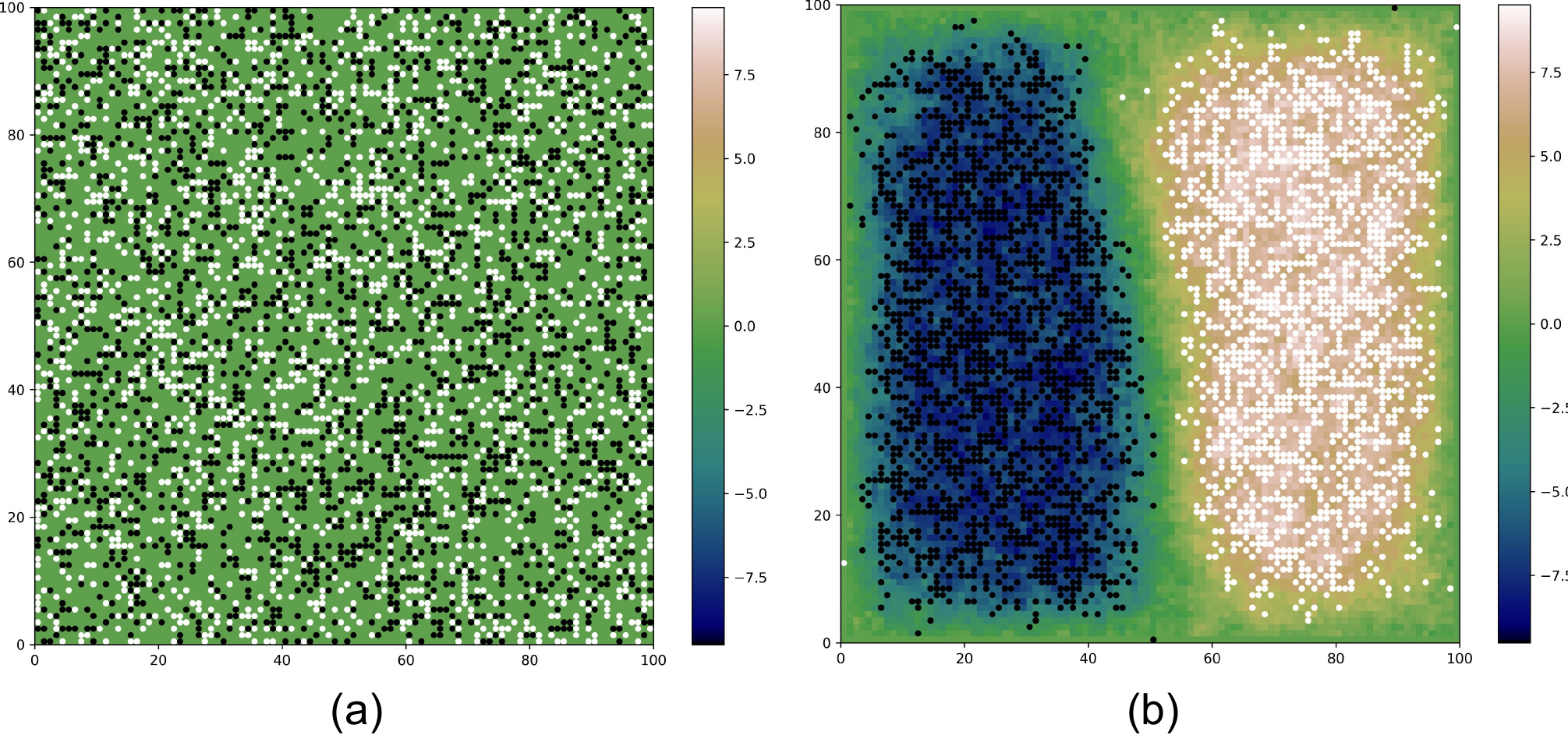}
  \caption{Phase separation between passive opposite particles arising from the field-mediated interactions. The particles are the white and black dots and the color map indicates the field. Hard walls prescribe a zero field on the boundary. (a) Initial state at $t=0$. (b) Complete separation of two liquid domains at $t=6000$. Parameters: $L=100$, $\rho=0.4$, $r=0.01$, $B=0.17$, $\phi_0=8$, $\mu=10$, $\tau=10^{-4}$, identical to those of fig.~\ref{fgr:LG} expect for the density and the system size.}
  \label{fgr:Phase_sep}
\end{figure}

We have also investigated the repulsion between opposite particles using simulation~1 with same physical parameters (fig.~\ref{fgr:Phase_sep}) . Starting from a random distribution of the two kinds of particles, we observe a phase separation between opposite liquid phases. This was expected since like particles attract and opposite particle repel. The phase separation is complete. Note how the liquid phases are compressible: their density $\rho>0.4$ is more than twice the equilibrium density $\rho_l\approx0.2$ obtained in fig~\ref{fgr:LG}.

\section{Retarded effects (passive particles)}

\subsection{Particles celerity}

Let us define the microscopic celerity of the particles as the ratio $c_p=\tau_f/\tau_p$, where $\tau_p$ is the time that a particle needs to diffuse on its own size, $a$, and $\tau_f$ is the time that the field needs to relax on that length scale. Restoring momentarily the dimensions, we have $\tau_p=a^2/(\kt\mu)$. From eq.~\eqref{eq:field_dyn}, a mode $q$ relaxes in a time $\tau(q)=\Gamma^{-1}/(r+cq^2)$. The  mode corresponding to an oscillation at the scale of the particles is the largest mode in the first Brillouin zone, i.e., $q_\text{max}=\pi/a$. It relaxes on a time $\tau_f=\Gamma^{-1}/(r+c\pi^2/a^2)\simeq a^2/(\pi^2 c\Gamma)$. The latter approximation holds since $q_\text{max}\gg1/\xi=\sqrt{r/c}$. We thus obtain $c_p=\kt\mu/(\pi^2c\Gamma)$. In dimensionless form this gives $c_p=\mu/\pi^2$. The threshold value above which the particles are effectively fast compared to the field (at their own scale) is therefore
\begin{align}
\mu_0=\pi^2\simeq10.
\end{align}

\subsection{Transient clusters}

Since the particles interact via the field, which may be slow relative to them, their effective interaction may exhibit time-dependent features due to retarded effects.

\begin{figure}
\centerline{\includegraphics[width=1\columnwidth]{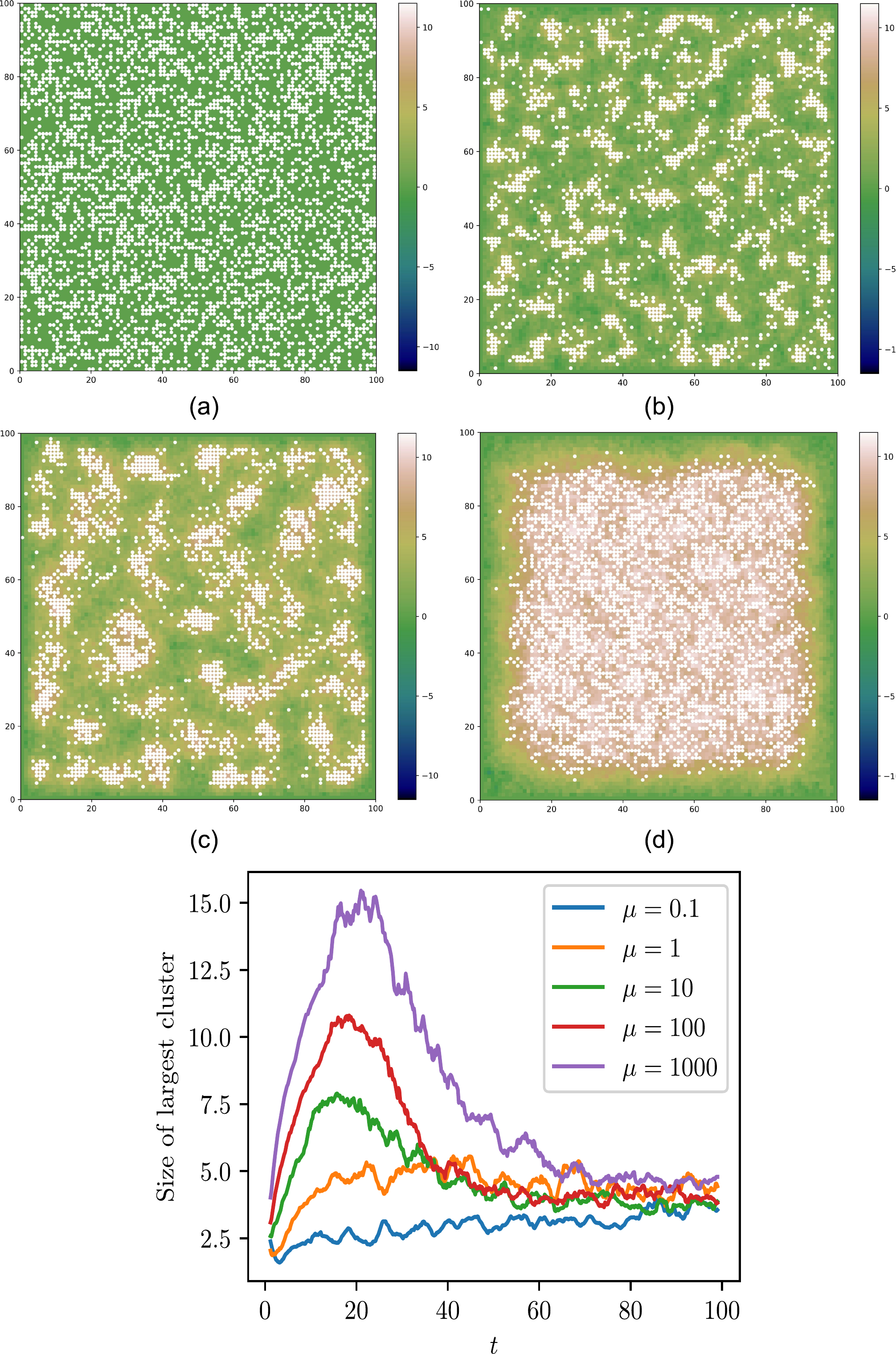}}
\caption{Field-mediated interactions between passive particles favoring a high field (white dot). In the initial state (a) the field is zero and the particles are  randomly distributed. The system develops transient clusters (b)-(c) that disappear in equilibrium (d). (a) $t=0$. (b) $t=4$. (c) $t=10$. (d) $t=100$. Parameters: $L=100$, $\rho=0.4$, $r=0.01$, $B=0.2$, $\phi_0=10$, $\mu=10$, $\tau=10^{-4}$. (e) Cluster size versus time for various values of $\mu$ (other parameters unchanged).
}
\label{fgr:1_}
\end{figure}

Using simulations 1, we have looked into the early stages of the  dynamics. With parameters in the same range as before, and $\mu=10$, we observe at the onset of the simulation the formation of transient clusters that grow for a while, then decrease in size, then disappear as the system reaches equilibrium (fig.~\ref{fgr:1_}a--d). They grow according to a multi-scale mechanism:   small clusters form due to the attraction of particles, then they grow either by attracting isolated particles or by merging with other clusters. This last process is slow because the large clusters diffuse slowly.

To measure the size of the clusters we proceed as follows. Since they  are denser than the accidental clusters formed randomly at $\rho=0.4$ (see fig.~\ref{fgr:1_}a), we first eliminate all the particles that have less than 6 neighbors among the eight possible nearest and next-nearest ones. We then identify the clusters using the nearest-neighbor relation~\cite{percolation_book}. We define the size of largest cluster as the square root of the number of sites in the largest cluster. Running the simulation for $\mu$ in the range 0.1--1000, we find that the clusters form only when $\mu\so\mu_0$, at the early stages of the simulation, and disappear when the system reaches equilibrium (fig.~\ref{fgr:1_}e). This shows that they are due to retarded effects. We observe also that the time  at which the clusters reach their maximum size and the time at which they disappear is essentially independent of $\mu$.

Although the dynamics of the clusters is difficult to model, it is possible to predict when the clusters will start to form. Recall that a mode $q$  relaxes with a rate $R(q)=r+q^2$, i.e., $R(q)\simeq q^2$ at scales smaller than the correlation length $\xi$. Take then a fixed particle that starts to deform an initially zero field at $t=0$. Since at time $t$ the modes with $R(q)\gg t^{-1}$ have reach their equilibrium deformation, while the modes $R(q)\ll t^{-1}$ still have essentially a zero amplitude, the characteristic width $w(t)$ of the deformation is such that $R(\pi/w)\approx t^{-1}$. This gives a diffusive growth $w\approx\pi\,t^{1/2}$. Now, since the typical inter-particle distance is $\rho^{-1/2}$, two particles will start to interact at a time $t_\text{int}$ such that $\rho^{-1/2}\approx\pi\,t_\text{int}^{1/2}$ (overlap of the deformations). This gives
\begin{align}
t_\text{int}\approx\frac1{\pi^2\rho},
\end{align}
which is independent of the parameters $r$, $B$, $\phi_0$ and $\mu$. Since for $\mu\ge\mu_0$ the particles are faster than the field, we expect the clusters to form after a few times $t_\text{int}$. For $\rho=0.4$, we obtain $t_\text{int}\approx0.25$ (well above the time resolution used). In the simulation, we find that clusters form at $t_0\simeq0.4$ for $\mu=1000$ and $\mu=100$ and at $t_0\simeq1$ for $\mu=10$, which confirms that $t_\text{int}$ is indeed the characteristic time of formation of the clusters.

\section{Randomly switching particles (active)}\label{sec:lumps}

\begin{figure}
\centering
  \includegraphics[width=1\columnwidth]{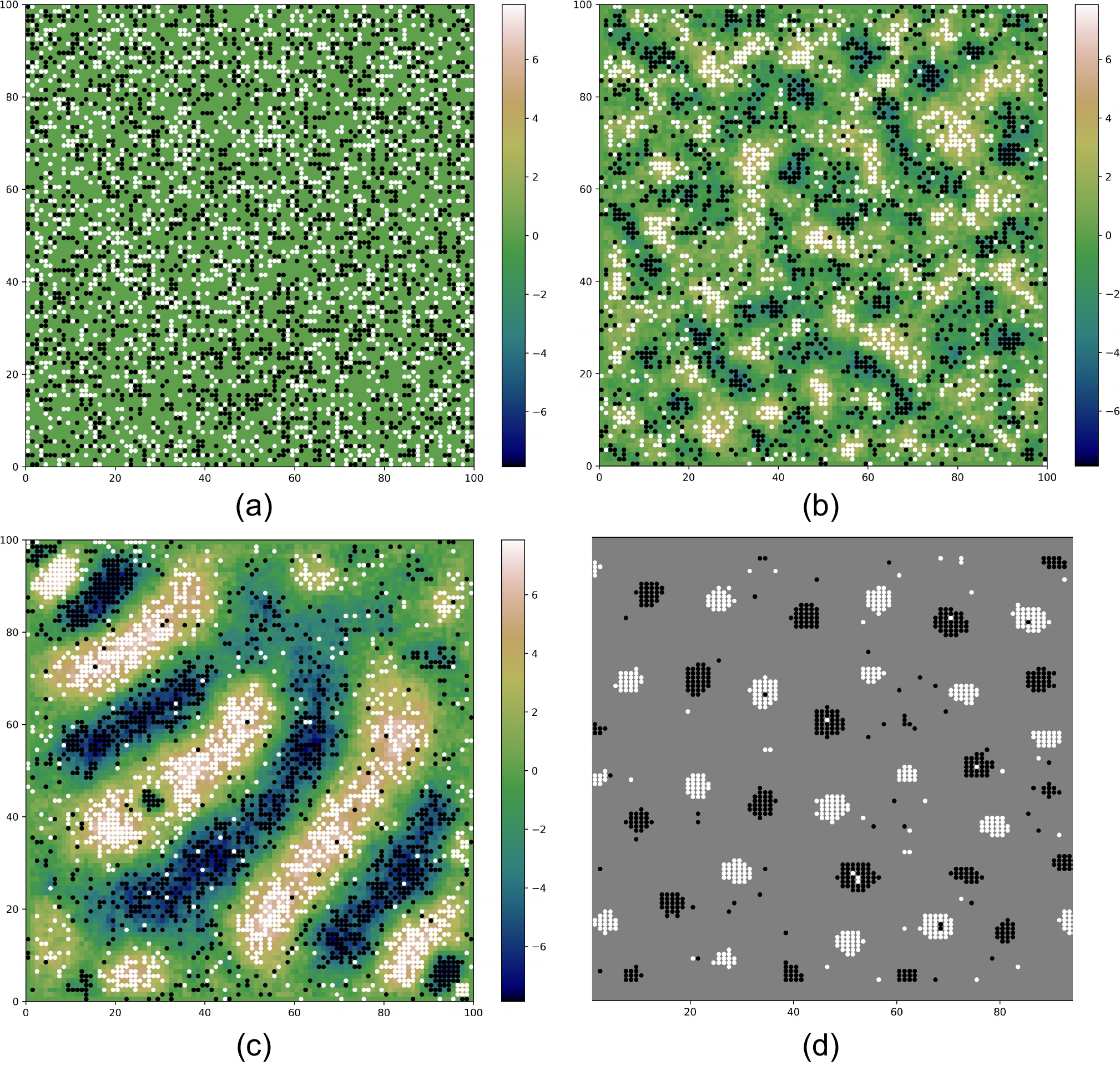}
  \caption{Field-mediated interactions between active particles randomly switching conformation (black $\leftrightarrow$ white dots). (a--c) Field-mediated interactions. Initial state (a): zero field and random mixture of white and black particles. The boundaries are hard walls prescribing a zero field. The systems develops alternate bands (c) in which the  particles either drift away to adjacent bands after switching, or form lumps that also drift away (black spot in the white band on the left). The field is high in the bands of white particle and low in the bands of black particles. (d) Corresponding direct pairwise interactions: dense clusters form that slowly merge. Particles having recently switched are visible inside the clusters. (a) $t=0$. (b) $t=10$. (c) $t=500$, $o=1.4$. (d) $t=15$, $o=3.9$. Parameters: $L=100$, $\rho=0.4$, $r=0.01$, $B=0.17$, $\phi_0=8$, $\mu=10$, $\alpha=0.1$, $\tau=10^{-4}$.}
  \label{fgr:2f}
\end{figure}

We now consider particles that switch randomly with a fixed rate $\alpha$ between the black ($S_k=-1$) and white ($S_k=1$) conformational states, as described in eqn~\eqref{eq:spin_flips}. The phase diagram and the nonequilibrium behavior of this active system were discussed in Ref.~\cite{Zakine:2018,Zakine:2020}. Here, we provide new results on the so-called ``lumps", which are specific recurring dynamical structures that appear within the bands that are formed by the system (fig.~\ref{fgr:2f}c).

The mechanism by which the bands form is as follows. Consider a system with an initial mixture of black and white particles (fig.~\ref{fgr:2f}a). As we have seen in the previous sections, the system will attempt to phase separate into two coexisting liquid phases of opposite states. During the phase separation process, many particles switch state and, as a result, black particles appear in the white region and vice versa. As like particles attract, clusters of identical particles form everywhere (fig.~\ref{fgr:2f}b). If the mobility of the particles is large enough (see below), these clusters percolate and form bands (fig.~\ref{fgr:2f}c), because this corresponds to a stationary state where no new aggregates form as the newly switching particles are individually expelled towards the adjacent bands of opposite state.

\begin{figure}
\centering
  \includegraphics[width=1\columnwidth]{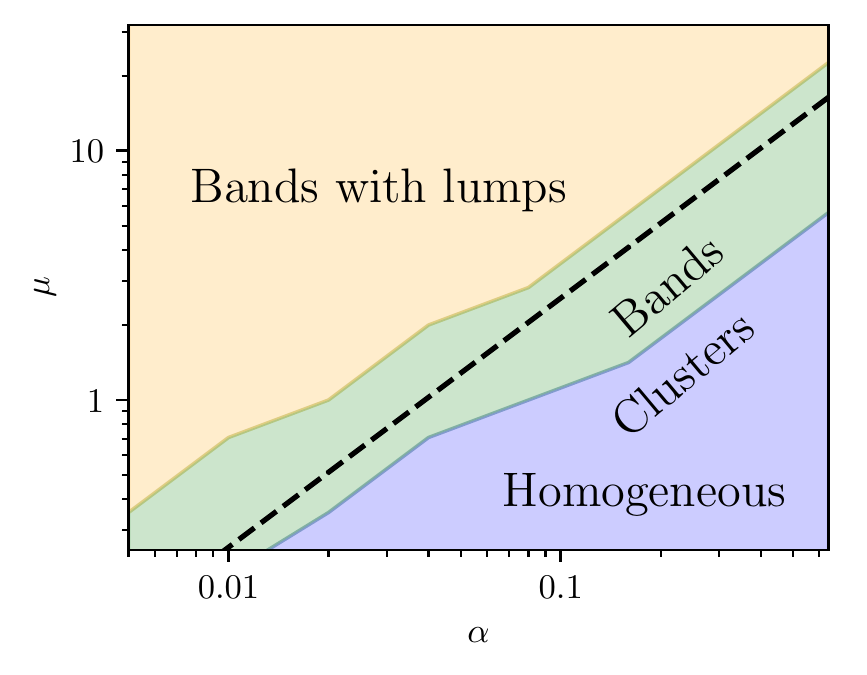}
  \caption{Phase diagram for the formation of lumps as a function of  particles switching rate $\alpha$ and mobility $\mu$, for $L=100$, $\rho=0.4$, $r=0.01$, $B=0.17$, $\phi_0=8$ and $\tau=10^{-4}$. (blue) Homogeneous system or small fluctuating clusters. (green) Narrow bands with no lumps. (cream) Large bands with recurrent lumps. The dashed line correspond to criterion \eqref{eq:lump_cri}.}
  \label{fgr:lump_criterion}
\end{figure}

Linear stability analyis~\cite{Zakine:2020} shows that the system forms clusters, or bands, when $\ell_d>\ell_0$, where
\begin{align}
\ell_d=\sqrt{\frac{\mu}{2\alpha}}
\end{align}
is a typical diffusion length of the particles between two flips, and $\ell_0^{-1}=B\phi_0\sqrt{\rho}-\sqrt{r+B\rho}$ depends on the system's parameters. Indeed, if the particles switch too rapidly ($\alpha$ large) or are too slow ($\mu$ small), the local phase separation does not have time to take place. With typically in our simulations $B=0.17$, $\phi_0=8$, $\rho=4$ and $r=0.01$, this gives $\alpha/\mu\lesssim0.2$ in agreement Fig.~\ref{fgr:lump_criterion}. The wavevector $k_d$ of the bands is given by the geometric mean~\cite{Zakine:2020}
\begin{align}
k_d=\left(\ell_d\tilde\xi\right)^{-1/2},
\end{align}
where $\tilde \xi=(r+B\rho)^{-1/2}$ is the correlation length of the field, renormalized by the presence of the particles.

In some regions of the phase diagram, black lumps regularly appear in the middle of the white bands (fig.~\ref{fgr:2f}c) and vice versa. These lumps diffuse, then drift to adjacent bands and merge with them. We have observed that the lumps tend to appear in the middle of the bands,  especially when the bands are large. We therefore propose the following mechanism. When the width of the bands is smaller than the correlation length, the newly switched particles feel the field gradient caused by the band structure and they are driven to the adjacent band before they can aggregate other particles. In this case, no lumps appear. On the contrary, if the bands are larger than the correlation length, the newly switched particles that appear in the middle of a band, feel no field gradient and no force, so they form new aggregates: these are the lumps. These lumps will eventually diffuse close to the side of the bands, feel the gradient, and get expelled. The criterion for lump formation, discussed above, is therefore $k_d\tilde \xi\lesssim1$, which leads to
\begin{align}
\frac\mu\alpha\gtrsim\frac2{r+B\rho}.
\label{eq:lump_cri}
\end{align}
As can be seen in fig.~\ref{fgr:lump_criterion}, this criterion works reasonably well (see the dashed line), which validates the above mechanism.

\section{Synchronously switching particles (active)}

Driving the system out of equilibrium can also be achieved  by forcing the particles to switch  state collectively (e.g., under a broad external signal). Here, we take particles that switch synchronously every $\Delta t$ (eqn~\eqref{eq:spin_synchro}).

\begin{figure}
\centering
  \includegraphics[width=1\columnwidth]{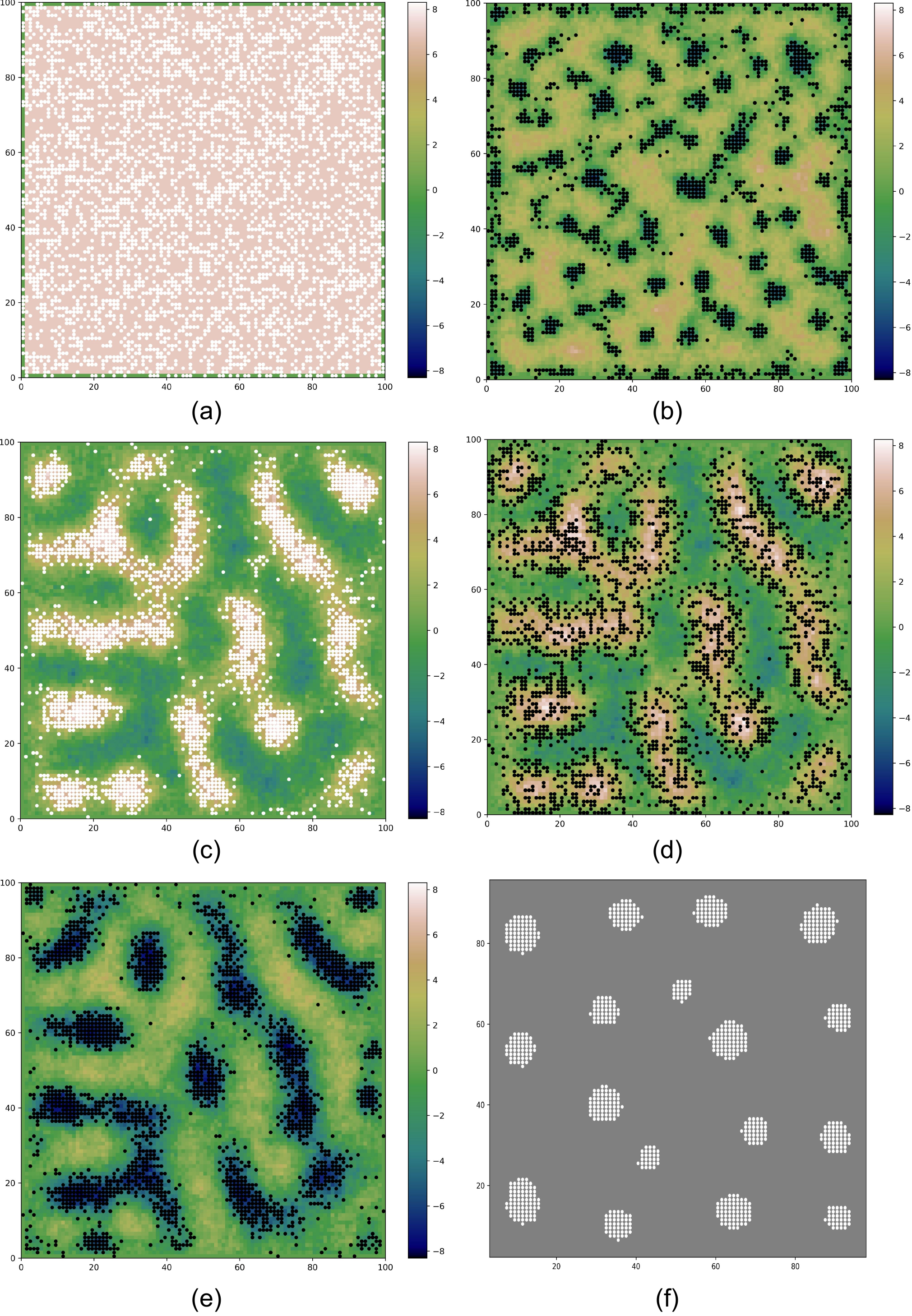}
  \caption{Field-mediated interactions between active particles switching conformation synchronously with period $\Delta t$ (all black $\leftrightarrow$ all white). (a-e) Field-mediated interactions.  Initial state (a): equilibrium field and white particles. The boundaries are hard walls prescribing a zero field. The particles switch to black at $t=0$ and start to form clusters (b). In the steady state, the system develops alternate bands, alternatively full and empty (c and e). When the particles are white (resp.\ black) they accumulate in the bands of high (resp.\ low) field. At each switch the (newly switched) particles move collectively from the occupied bands to the empty adjacent ones (c $\leftrightarrow$ e). Image (d) shows the particles starting to move towards the adjacent empty bands just after a switch occurs in image (c). (f) Corresponding direct pairwise interactions: dense clusters form that slowly merge and simply change color at each switch (white $\leftrightarrow$ black). (a) $t=0$. (b) $t=11$. (c) $t=481$. (d) $t=481.2$. (e) $t=491$. (f) $t=15$. Parameters: $\rho=0.4$, $r=0.01$, $B=0.17$, $\phi_0=8$, $\mu=10$, $\Delta t=10$, $\tau=10^{-4}$.}
  \label{fgr:1s}
\end{figure}

Let us consider first the case where the particles are all initially in the same state (fig.~\ref{fgr:1s}). They thus become alternately all white ($S_k=1$) and all black ($S_k=-1$). When the particles are fast compared to the field, the systems  forms alternate bands of low and high field (fig.~\ref{fgr:1s}c) as in the case of randomly switching particles. These bands are fixed in space and the particles collectively jump every $\Delta t$ from the regions of high field, that they populate when they are white, to the regions of low field, that they populate when they are black  (fig.~\ref{fgr:1s}c--e). 

The mechanism of formation of these bands is the following. Initially, the particles are all white and randomly distributed (fig.~\ref{fgr:1s}a). At $t=0^+$ they become all black and start to aggregate. Underneath the forming black aggregates, regions of low field appear. When at $t=\Delta t$ the particles all become white again, they are expelled from these now unfavorable low field regions. Because the field is relatively slow, the low-field regions do not have the time to relax their footprints, and the white aggregates that are now forming are forced to surround them. Then, at the next switch, the newly black particles recolonize the low-field regions, forming aggregates that merge in the valleys surrounding the footprints of the high field regions. Over time, the low and high field regions form alternate band-like structures.

Decreasing the mobility $\mu$ of the particles, while keeping the other parameters as in fig.~\ref{fgr:1s}, leads to the disappearance of the bands: for $\mu=5$ they become somewhat blurred and for $\mu\le2.5$ we observe only irregular clusters which are not significantly altered by the switching of the particles. Also, if we replace the multibody field-mediated interactions by its pairwise component $F_\text{pw}$ only, no bands are formed: the system simply forms dense clusters that collectively switch state each $\Delta t$, with no incidence on their shapes and organisation (fig.~\ref{fgr:1s}f). These two observations show that the bands displayed in fig.~\ref{fgr:1s}c result from the interplay of multibody interactions and retardation effects.

\begin{figure}
\centering
  \includegraphics[width=1\columnwidth]{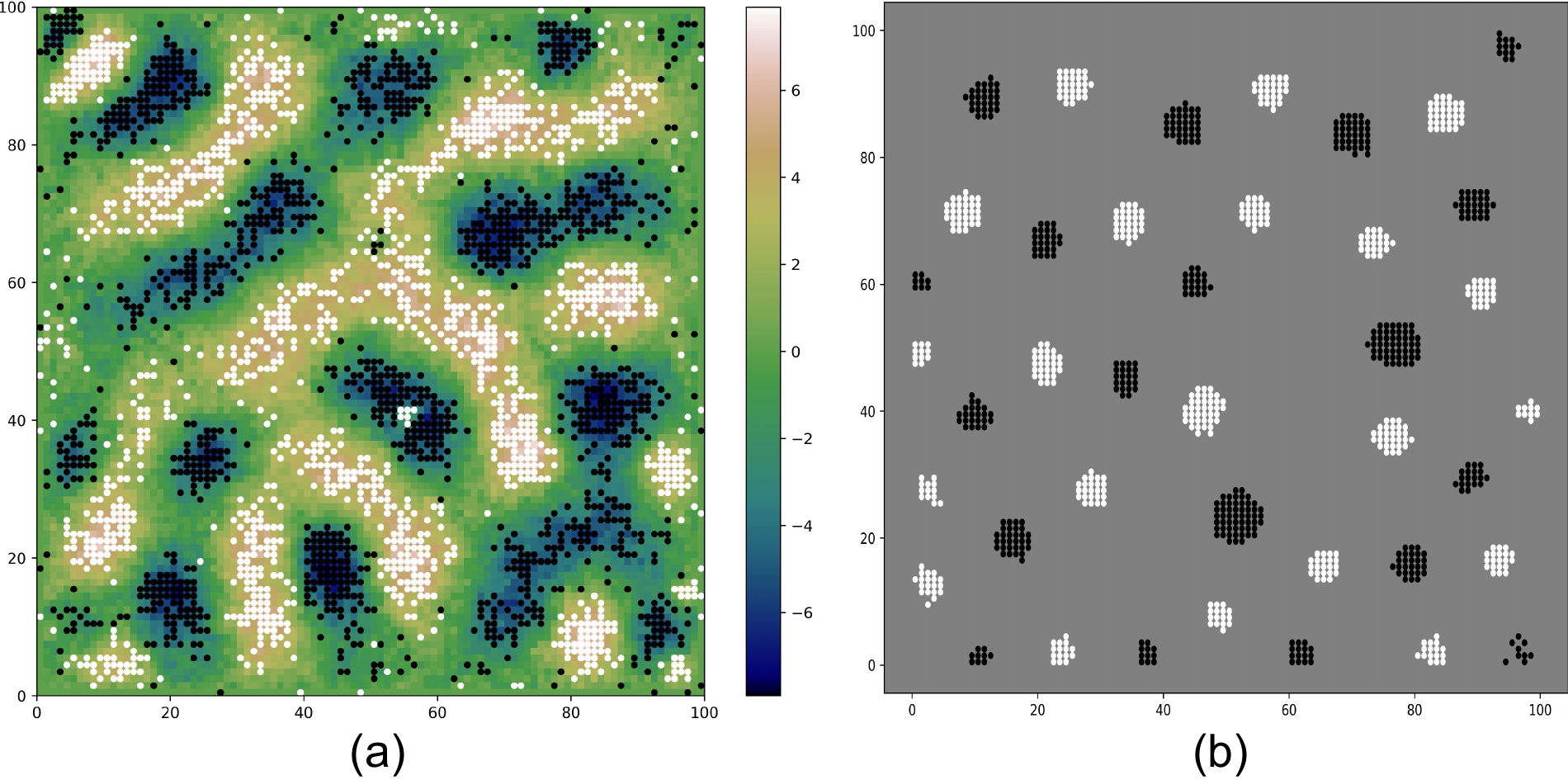}
  \caption{Field-mediated interactions between active particles switching  synchronously with period $\Delta t$ (all black $\leftrightarrow$ all white). The conditions and parameters are the same as in Fig.~\ref{fgr:2f}, except that the switches are synchronous with $\Delta t=10$ instead of been random. (a) Field-mediated interactions. (b) Corresponding direct pairwise interactions.  Despite this fundamental difference, the steady state in the field-mediated case (a) is very similar, lumps included (compare with Fig.~\ref{fgr:2f}c).}
  \label{fgr:2s}
\end{figure}

We now consider the case where the initial state consists of a random mixture of white and black particles. As before, we force the particles to change state simultaneously every $\Delta t$. It is interesting to note that alternating bands are also formed in this case (fig.~\ref{fgr:2s}). In fact, the steady state resembles that of fig.~\ref{fgr:2f} (randomly switching particles), including the presence of lumps. The mechanism of band formation is however quite different. Here the bands are formed from the mechanism described just above (in the present section). The principal difference is that the high-field and low-field regions are both populated, and that the corresponding white and black populations simply exchange their positions every $\Delta t$. Lumps are formed for the same reason as in section~\ref{sec:lumps}: in a wide band, the recently switched particles experiencing an unfavorable field are not expelled (since they do not feel any gradient), but start to aggregate by altering the local field, forming clusters.

\section{Summary and discussion}

Particles coupled to a correlated field undergo field-mediated interactions. Although the elasticity of the field, its dynamics, and the way the particles are coupled to the field are specific, basic properties and non-equilibrium trends can be examined using the simplest model: pointlike particles coupled quadratically to a Gaussian field with overdamped dynamics~\cite{Netz:1997,Dommersnes:1999,Zakine:2018}. We therefore studied this system in detail. Since one can calculate exactly the multibody interaction energy in an assembly of an arbitrary number of particles, we could systematically compare the behavior of the real system with that which would result from including only the pairwise component of the interaction. We found that in a dense assembly of particules, the energy per particle is much lower than what would result from considering pairwise interactions alone (see Fig.~\ref{fgr:Multibody_vs_pairwise}). This property is most likely generic for field-mediated interaction, as it originates from the fact that in a dense assembly the field is everywhere almost equal to the value that satisfies the particles, so that adding or removing one particle is almost costless (in the sense that bringing one particle in a dense assembly essentially relaxes its self-energy).

We have studied the gas-liquid phase transition in the system and obtained several results that we believe should be generic for field-mediated interactions, as they are consequences of the effective weakness of the multibody interactions in a dense system.
First, the liquid phase is relatively sparse and highly compressible (see Fig.~\ref{fgr:LG}, and compare with Fig.~\ref{fgr:Phase_sep} where the density of the liquid is twice as large due to extra pressure). Second, the mean-field density functional approach works well in the liquid phase due to the relatively homogeneous value of the field.

Another characteristic of the field-mediated interaction, generic but not systematic, because it depends on the relative celerity of the field and the particles, is the possibility of observing transient structures due to retardation effects.
In a system of particles with the density of the liquid phase, we examined what happens if we start from an initial condition where the field is vanishing, while using fast particles. In agreement with what has been said above, the field not being in equilibrium, the mediated interactions are much more important. Since the particles move rapidly, this produces dense clusters (see Fig.~\ref{fgr:1_}). These clusters are transient and disperse when the field reaches equilibrium, resulting in the equilibrium liquid phase. High transient mediated-interactions were also reported in Refs.~\cite{Fournier:2013,Fournier:2014}.

Let us discuss whether or not these retardation effect can be observed in actual systems. For membrane proteins no such effects should occur. Indeed, with $a\simeq\SI{3}{nm}$ and $D\simeq\SI{10}{\mu m^2/s}$ (see, e.g., Ref.~\cite{Weiss:2013}), we obtain $\tau_p=(2a)^2/D\simeq\SI{3.6}{\mu s}$. With $\kappa\simeq\SI{e-19}{J}$ and $\eta\simeq\SI{e-3}{J.s.m^{-3}}$, we get $\tau_f=4\eta/(\kappa q^3)\simeq\SI{0.3}{ns}$ for $q\simeq\pi/a$, giving a ratio $c_p\approx10^{-4}\ll1$. For macroscopic objects floating at a gas-liquid interface, we do not expect these effects either, due to the smallness of the diffusion coefficient, which is proportional to the inverse particle size. To observe the retardation effects discussed above, it would be necessary to slow down the field dynamics. This might be possible for very small colloids trapped at an interface between matched fluids with very low surface tension, or for small colloids in a binary mixture very close to the critical point, where critical slowing down occurs.

We also reviewed the bands of liquid phases composed of particles of opposite states~\cite{Zakine:2018,Zakine:2020}, and the associated lumps, which appear in active systems where particles change state randomly. In particular, we have obtained a simple criterion for the generation of lumps. We have shown that these patterns disappear when the multibody interaction between particles is replaced by its single pairwise component. They are thus the result of the combination of non-equilibrium activity and multibody interactions. These patterns are not due to retardation effects, they are caused by the microscopic breaking of the coarsening that is caused by the random switch of the particles. We therefore expect to find them generically in real systems. Indeed, the criterion for their appearance is $\ell_d$ (typical diffusion length between two particles switch state) larger than some threshold (independent of the dynamics), which can be satisfied even for slow particles if the switching rate is low enough.

We have also shown the possibility of interesting retardation effects: band-like structures in a system composed of only one type of particles changing state simultaneously. We have uncovered the mechanism responsible for these patterns and also shown that it is the result of a combination of non-equilibrium activity and multi-body interactions. The particles create imprints in the field, that do not have time to relax, and they periodically migrate between them. Interestingly, a system composed of two types of synchronously switching particles also exhibits bands and lumps, not because of the mechanism that is at play for randomly switching particles, but because of the imprinting mechanism. As with the appearance of transient clusters discussed above, this effect should only be observable in real systems if the field dynamics is slowed down.

\appendix
\section*{Appendix}

\section{Numerical simulations}
\label{apx:numerical_simulations}

\subsubsection*{Simulation~$1$: field-mediated interactions}

We simulate the system comprizing the field and the particles on a two dimensional square lattice of size $L$ as in Ref.~\cite{Zakine:2018}, except that we take with impenetrable walls and set $\phi=0$ on the boundary  (unless otherwise specified).
The field $\phi_{i,j}$ is defined on each lattice site and takes continuous values, while the particles hop from site to site. 
The discretized Hamiltonian is given by
\begin{align}
H=\sum_{i,j}\left[\frac r 2 \phi_{i,j}^2+
\frac12\left(\bm \nabla\phi\right)^2_{i,j}
\right]+\sum_{k=1}^N \frac B 2 \left(\phi_{i_k,j_k}-S_k\phi_0 \right)^2,
\end{align}
with the standard forward gradient
\begin{align}
\left(\bm\nabla h\right)_{i,j}&=\left(h_{i,j+1}-h_{i,j},~h_{i+1,j}-h_{i,j}\right)^t.
\end{align}

Every time step $\tau$, we update the field according to the discretized Langevin equation $\phi_{i,j}(t+\tau)-\phi_{i,j}(t)=-\tau\partial H/\partial h_{i,j}+\xi_{i,j}(t)$ where $\xi_{i,j}(t)$ is a random Gaussian variable of mean zero and variance $2\tau$. Calculating explicitly the partial derivative of $H$ with respect of $\phi_{i,j}$ yields
\begin{align}
\phi_{i,j}(t+\tau)&=\phi_{i,j}(t)
-\tau\left[
r\phi_{i,j}(t)-
(\bm\nabla^2\phi)_{i,j}(t)
\right]
\nonumber\\
&-\tau\sum_{k=1}^N B\left[\phi_{i,j}(t)-S_k(t)\phi_0 \right]\delta_{i,i_k}\delta_{j,j_k}+\xi_{i,j}(t).
\label{evolution_phiij}
\end{align}
with the discrete Laplacian
\begin{align}
(\bm \nabla^2\phi)_{i,j}&=
\phi_{i,j+1}+\phi_{i,j-1}+\phi_{i-1,j}+\phi_{i+1,j}-4\phi_{i,j}.
\end{align}

For the update of the particles positions, we choose $N$ times at random a particle among the $N$ particles and let it hop to a neighbouring site, flips or do nothing. To this aim, we calculate the probabilities $P_r$, $P_u$, $P_l$, $P_d$, $P_f$ for a particle to move right ($j\to j+1$), up ($i\to i-1$), left ($j\to j-1$), down ($i\to i+1$) or to flip, respectively.
When a particle moves from $(i,j)$ to $(i',j')$ the energy variation is
\begin{align}
\Delta H_{r,u,l,d}=
\frac{B}{2}(\phi_{i'_k,j'_k}-\phi_{i_k,j_k})(\phi_{i'_k,j'_k}+\phi_{i_k,j_k}-2 S_k\phi_0).
\end{align}
Choosing symmetric equilibrium Boltzmann hoping rates (satisfying detailed balance), we take
\begin{align}
P_{r,u,l,d}=\mu\tau\,e^{-\Delta H_{r,u,l,d}/2}.
\end{align}

The flip probabilities depend on our choice of the dynamics of the state variables. For time-independent states, we simply take $P_f=0$. For random flips at fixed rates, that break detailed balance, we take
\begin{align}
P_f=
\alpha\tau.
\end{align}
Finally, for synchronized flips, we simply change each $S_k$ into $-S_k$ every $\Delta t$, which agains break detailed balance.

\textit{Force acting on a particle.}---The force acting on particle $k$ is given from eqn~\eqref{eq:Hamiltonian} by
\begin{align}
\bm f_k=-\frac{\partial \mathcal H}{\partial\bm x_k}
=-B\left[\phi(\bm x_k)-S_k\phi_0\right]\bm\nabla\phi(\bm x_k).
\end{align}
Using the discretized Hamiltonian, the $x$-component of the force acting on particle $k$ is therefore
\begin{align}
f = -\frac B2(\phi_{i_k,j_k}-\phi_0)(\phi_{i_k,j_k+1} - \phi_{i_k,j_k-1}),
\label{eq:discretized_force}
\end{align}
with the symmetrized discrete gradient.

\subsubsection*{Simulation~$2$: Direct pairwise interactions}

Simulation~$2$ is akin to simulation~$1$, except that there is no field $\phi_{i,j}$ and that $\Delta H_{r,u,l,d}$ is computed from the variation, when the particle moves from $(i,j)$ to $(i', j')$, of the sum of all its interactions $F_\text{pw}$ with the other particles  (see eq.~\eqref{eq:Fpw}, Appendix~\ref{apx:field-mediated_interactions}).

\onecolumngrid
\section{Field-mediated interactions}
\label{apx:field-mediated_interactions}

The total field-mediated interaction energy between $N$ particles can be calculated, for fixed positions of the particles, from the partition function $Z=\int\!\mathcal{D}[\phi]\, e^{-\mathcal H}$, where $\mathcal H$ is given by \eqref{eq:Hamiltonian} in dimensionless form. Although the Hamiltonian is Gaussian, it is not translationally invariant, so it is useful to perform the following transformation:
\begin{align}
Z&=\int\!\left(\prod_{k=1}^Nd\lambda_k\right)
\exp\!\left(-\frac1{2B}\sum_k\lambda_k^2-i\sum_k\lambda_kS_k\phi_0\right)\,Z_\phi,\\
Z_\phi&=\int\!\mathcal D[\phi]\,\exp\!\left(-\int_{\bm x,\bm y}\frac12\phi(\bm x)H(\bm x,\bm y)\phi(\bm y)
+\int_{\bm x}i\phi(\bm x)\sum_k\lambda_k \delta(\bm x-\bm x_k)\right),
\end{align}
where $\int_{\bm x,\bm y}\equiv\int d^2x\,d^2y$, and $H(\bm x,\bm y)=(r-c\bm\nabla^2)\delta(\bm x-\bm y)$. Performing the Gaussian field integral gives
\begin{align}
Z_\phi=\exp\!\left[-\frac12\int_{\bm x,\bm y}\left(\sum_k\lambda_k \delta(\bm x-\bm x_k)\right)G(\bm x-\bm y)\left(\sum_{k'}\lambda_{k'} \delta(\bm y-\bm x_{k'})\right)\right]
\end{align}
where $G(\bm x-\bm y)$ is the correlation function, i.e.,
\begin{align}
Z=\int\!\left(\prod_{k=1}^Nd\lambda_k\right)\,
\exp\!\left(-\frac12\mathsf{\Lambda}\,\mathsf{A}\,\mathsf{\Lambda}^t-i\,\mathsf{\Lambda}\,\mathsf{S}^t\phi_0^N\right),
\end{align}
where $\mathsf{\Lambda}=(\lambda_1\ldots\lambda_N)$, $\mathsf{S}=(S_1\ldots S_N)$ and the $N\times N$ matrix $A$ has the elements:
\begin{align}
A_{ij}=
\begin{cases}
B^{-1}+G_0 & (i=j),\cr
G(\bm x_i-\bm x_j) & (i\ne j).
\end{cases}
\end{align}

The free energy of the system is thus 
\begin{align}
F=-\ln Z=\frac12\phi_0^2
\mathsf S{\mathsf A}^{-1}\mathsf S^t
+\ldots
\end{align}
where the ellipsis represents the fluctuation-induced, Casimir-like, interaction $\frac12\ln(\det\mathsf A)$. If $\phi_0$ is large enough, which is the case with the parameters used in this work, this contribution is negligible.

The mediated \textit{interaction} energy is given by $F_\text{int}=F-F_\text{self}$, where $F_\text{self}=Nf_\text{self}$ is the self-energy of the particles. Placing the particles infinitely far from one another, from $\mathsf A_\infty=(B^{-1}+G_0)\mathsf I_N$ where $\mathsf I_N$ is the identity matrix, we obtain
\begin{align}
f_\text{self}=\frac12\,\frac{\phi_0^2}{B^{-1}+G_0}.
\label{eq:f_self}
\end{align}
The mediated interaction energy is  therefore given by
\begin{align}
\label{mediated_interaction}
F_\text{int}=\frac12\phi_0^2(S_1\ldots S_N){\mathsf A}^{-1}(S_1\ldots S_N)^t-
\frac12\,\frac{N\phi_0^2}{B^{-1}+G_0}.
\end{align}
For two particles, this yields the pairwise interaction energy:
\begin{align}
F_\text{pw}(R)=-\phi_0^2s_1s_2\,\frac{G(R)}{\left(B^{-1}+G_0\right)
\left[B^{-1}+G_0+s_1s_2G(R)\right]}.
\label{eq:Fpw}
\end{align}

The correlation function $G$ is solution of $(r-\bm\nabla^2)G(\bm x)=\delta(\bm x)$, hence it is given by
\begin{align}
G(R)&=\frac1{2\pi}\text{K}_0(R\sqrt{r}),\\
G(0)=G_0&=
\int_0^\pi\frac{d^2q}{(2\pi)^2}\,\frac1{r+q^2}
=\frac1{4\pi}\ln\!\left(1+\frac{\pi^2}{r}\right)
\nonumber\\
&\simeq\frac1{2\pi}\ln\!\left(\frac\pi{\sqrt{r}}\right),
\end{align}
where $\text{K}_0$ is the Bessel function. We have regularized $G_0$ by using a high-wavevector cutoff $q_\text{max}=\pi/a$, with $a=1$ in our dimensionless units.
The results depend on the precise choice of the cutoff, but only weakly thanks to the logarithm in the formula giving $G_0$.

\textit{Case of a linear particle-field coupling.}--- 
In soft matter, particles are generally subject to fixed, or strongly enforced, boundary conditions (contact angle, deformations, etc.), so they are subject to quadratic or nonlinear couplings with the surrounding field. As a consequence, they experience multibody (and Casimir) interactions. We show here that using  a linear field-particle coupling would result in missing the multibody and Casimir contributions. If we change the Hamiltonian into
\begin{align}
\mathcal H_\text{lin} = \int d^2x\left[\frac r2\phi^2+\frac12c(\bm\nabla\phi)^2\right]
-\sum_{k=1}^NBS_k\phi(\bm x_k),
\label{eq:Hamiltonian_lin}
\end{align}
which is phenomenologically similar to $\mathcal H$, the partition function is given by
\begin{align}
Z&=\int\!\mathcal D\phi\,\exp\!\left(-\int_{\bm x,\bm y}\frac12\phi(\bm x)H(\bm x,\bm y)\phi(\bm y)
+B\int_{\bm x}\phi(\bm x)\sum_kS_k\delta(\bm x-\bm x_k)\right),
\nonumber\\
&=Z_0\exp\!\left[\frac{B^2}2\int_{\bm x,\bm y}\left(\sum_kS_k \delta(\bm x-\bm x_k)\right)G(\bm x-\bm y)\left(\sum_{k'}S_{k'} \delta(\bm y-\bm x_{k'})\right)\right]
\nonumber\\
&=Z_0\exp\left(
\frac{B^2}2\sum_{k,k'} S_k S_{k'} G(\bm x_k-{\bm x}_{k'})
\right).
\end{align}
This yields the total field-mediated interaction
\begin{align}
F'_\text{int}=-\ln Z=F_0-\frac{B^2}2\sum_{k,k'} S_k S_{k'} G(\bm x_k-{\bm x}_{k'}),
\end{align}
which is purely pairwise. There is also no fluctuation-induced Casimir interaction since $F_0$ is independent of the positions of the particles.

\bigskip
\twocolumngrid

\section*{Acknowledgements}

I thank F. van Wijland for valuable discussions throughout the development of this work.


%

\end{document}